# Peculiarities of Ehrenfest equation for solids strained by uniaxial or hydrostatic pressure

Ph.B. Moin

*Research institutes "Synthesis"*

*82300, Truskavetska 125, Boryslav, Ukraine*

e-mail ph.moin@mail.ru

## Abstract

The Ehrenfest equation is derived rigorously for the case when an "effective" volume $V^* = AV$ of the strained solids is a continuous function of the temperature. The Ehrenfest equation for strained solids is generalized to an arbitrary temperature. Far from 0 K (in contrast to the situation near 0 K) the phase transition temperature derivative with respect to uniaxial pressure depends on the temperature expansion coefficient of an " effective" volume. For Rochelle salt strained by a uniaxial pressure at room temperature the predominant contribution into this dependence is made by the coefficient A. At the second order phase transition under uniaxial pressure the "effective" (but not true) volume is the continuous function of the temperature. The Ehrenfest equation for solids strained by hydrostatic pressure can be presented as a sum of three linear parts proportional to the crystal linear thermal expansion coefficients along the axes *a, b, c*. These parts are equal to the second order transition temperature derivative with respect to hydrostatic (but not uniaxial) pressure,

## Introduction

At the second order phase transitions the specific volume and entropy are continuous functions of temperature, while the heat capacity, volume thermal expansion coefficient, and compressibility have jumps. The superconducting phase transition in superconducting materials (including synthetic crystals, see [1-3]) at low temperature, ferroelectric transitions in triglycine sulfate at 322 K and in Rochelle salt at 255 K and at 297 K are the typical examples of the second order phase transitions, see [4-5]. Dependence of the second order phase transition temperature on hydrostatic pressure is described by the well-known Ehrenfest equation. In solids, the second order phase transition temperature can be changed not only by the hydrostatic pressure, but also by various non-hydrostatic stresses.

In [6] the Ehrenfest equation was rigorously generalized to the solids with arbitrary homogeneous elastic deformations accompanied by a change of the solids volume, in particular, to the solids strained by the uniaxial, biaxial, or triaxial pressure. For Ehrenfest equation generalization, the introduced in [7] concept of the "effective" volume $V^* = AV$ of the strained solids was used. It has been shown that for non-hydrostatic pressure the "effective" volume plays the role of the true one. The thermodynamic equations for the strained solids are obtained from the usual ones by a simple replacing V with V*. The proportionality coefficient A does not depend on the strain value and is determined by the elastic compliances of the solid and by the stress ratios. In [6] the exact Ehrenfest equation for solids strained by the non-hydrostatic pressure was derived for low temperatures (i.e. for superconductors). This derivation has been made at the assumption that the proportionality coefficient A does not depend on temperature. But for the second order phase transitions at higher temperatures it is necessary to take into account this dependence.

The aim of this paper is a rigorous generalization of the Ehrenfest equation for solids strained by the non-hydrostatic (in particular uniaxial) pressure to an arbitrary temperature. In addition, in this paper the possibility to present the second order phase transition temperature derivative with respect to hydrostatic pressure as a sum of three derivatives with respect to uniaxial pressures is considered.

1. The proportionality coefficients in the expression for the "effective" volume $V^* = AV$ of the solids strained by a uniaxial pressures, applied along the axes $a, b, c$   $A_a$, $A_b$, $A_c$ are equal to

$$A_a = S_{11}/(S_{11}+S_{12}+S_{13}),$$
$$A_b = S_{22}/(S_{21}+S_{22}+S_{23}), \qquad (1)$$
$$A_c = S_{33}/(S_{31}+S_{32}+S_{33}),$$

where $S_{ik}$ are the components of the elastic compliances tensor.
All $A_i$ are larger than 1. For hydrostatic pressure $A = 1$, (see [7]).

According to the Ehrenfest equation the derivative of the transition temperature with respect to hydrostatic pressure is the function of the volume specific thermal expansion coefficients $\alpha_1$ and $\alpha_2$ of the two equilibrium phases.

$$dT/dP = T\Delta(\partial V/\partial T)/\Delta C_P = TV(\alpha_1 - \alpha_2)/(C_{P1} - C_{P2}) \qquad (2)$$

The Ehrenfest equation for a non-hydrostatic pressure, in particular, for uniaxial one is obtained from Eq.(2) by simple replacing V with V*:

$$dT/dp_i = T\Delta(\partial V^*/\partial T)/\Delta C_P = T\Delta[\partial(AV)/\partial T]/\Delta C_P \qquad (3)$$

In [6], assuming that $\Delta V = 0$ (i.e. $V_1 = V_2 = V$) under hydrostatic and non-hydrostatic pressure and also that the coefficient A does not depend on temperature, the exact Eq.(3) has the form:

$$dT/dp_i = TV\Delta(A\alpha)/\Delta C_P = TV(A_1\alpha_1 - A_2\alpha_2)/(C_{P1} - C_{P2}) \qquad (4)$$

Thinking formally, using the concept of the "effective" volume [7], at the second order phase transition under non-hydrostatic pressure it is the "effective" volume jump that should be equal to zero: $\Delta V^* = \Delta(AV) = A_1V_1 - A_2V_2 = 0$.
In this case, since the two equilibrium phases have different elastic properties, in particular, the compressibilities, the coefficient A has a jump (see [6]), their partial volumes do not coincide, and the exact Eq.(3) has the form:

$$dT/dp_i = T(V_1A_1\alpha_1 - V_2A_2\alpha_2)/(C_{P1} - C_{P2}) \qquad (5)$$

Let us note that Eq. (5) is suitable for both cases: when $\Delta V = 0$ and when $\Delta V^* = 0$. Below we prove rigorously that at the second order phase transition under uniaxial pressure the "effective" (but not the true) volume is the continuous function of temperature.

2. The assumption about the independence of the coefficient A of temperature is correct at low temperatures (i.e. for superconductors), because the moduli of elasticity (and elastic compliances) approach the absolute zero with a zero slope [8]. At higher temperatures it is necessary to take into account the dependence of the coefficients $A_1$ and $A_2$ on T. In this case the Ehrenfest Eq.(3) obtains the form:

$$dT/dp_i = T\Delta(A\partial V/\partial T + V\partial A/\partial T)/\Delta C_P \qquad (6)$$

Introducing the designation of the volume thermal expansion coefficient $\alpha_V = (\partial V/\partial T)/V$; the designation of the A temperature coefficient $\alpha_A = (\partial A/\partial T)/A$; $\alpha = \alpha_V + \alpha_A$ we obtain the Ehrenfest Eq. (3) in the form, formally coinciding with Eq. (5). Now Eq.(5) is equally suitable both for hydrostatic and non-hydrostatic pressure, in particular, for a uniaxial one at any temperature. We must bear in mind, that $\alpha_1$ and $\alpha_2$ reflect the temperature dependence not only of the partial volumes $V_1$ and $V_2$, but also of the coefficients $A_1$ and $A_2$ of the two equilibrium phases.

3. Let us estimate the relative contributions of $\alpha_V$ and $\alpha_A$ in $\alpha$ for Rochelle salt in the orthorhombic modification at room temperature, for which the data for the elastic compliances and their temperature coefficients are available [8-9].

Table 1 Elastic compliances and their
temperature coefficients for Rochelle salt
at room temperature [9]

| $S_{ik}$ | $10^{-11}$ m²/N | $(\partial S_{ik}/\partial T)/S_{ik}$ $10^{-5}$K$^{-1}$ |
|---|---|---|
| $S_{11}$ | 5.16 | 123 |
| $S_{22}$ | 3.47 | 133 |
| $S_{33}$ | 3.33 | 89 |
| $S_{12}$ | - 1.58 | 524 |
| $S_{13}$ | - 1.70 | 271 |
| $S_{23}$ | - 1.0 | -1020 |

The coefficient $\alpha_A$ was evaluated in the following manner: the A values at room temperature were found from Eqs (1); then $\partial A/\partial T$ were calculated by differentiating Eqs (1) and using the data from Table 1. The values of $\alpha_A$ were found from the formula $\alpha_A = (\partial A/\partial T)/A$. The value of $\alpha_V$ is found as the sum of the linear thermal expansion coefficients $\alpha_a$, $\alpha_b$, $\alpha_c$ along axes *a, b, c* taken from Fig.3 Ref.[10]:

$$\alpha_V = \alpha_a + \alpha_b + \alpha_c \qquad (7)$$

Table 2 Temperature coefficients
and experimental $dT/dp_i$ data for 23.4 °C
phase transition in Rochelle salt.

|  | $A_a$ | $A_b$ | $A_c$ |  |
|---|---|---|---|---|
| $A_i$ | 2.74 | 3.90 | 5.29 |  |
| $\alpha_A$ $10^{-5}$K$^{-1}$ | 463 | -601 | -1269 |  |
| $\alpha_i$ $10^{-5}$K$^{-1}$ | 6.2 | 4.2 | 4.5 | [10] |
| $A\alpha_A 10^{-3}$K$^{-1}$ | 12.7 | - 23.4 | -67.1 |  |
| $dT/p_i$ K/kbar | 35.6 | - 18 | - 6.8 | [9] |
| $dT/p_i$ K/kbar | 35 | - 16 | -8* | [11] |

* The value calculated from hydrostatic data [11]

Thus $\alpha_V = 14.9 \cdot 10^{-5} K^{-1}$. The comparison of the $\alpha_A$ and $\alpha_V$ values reveals that $\alpha_V$ value is no more than 3.2 per cent of the absolute value of α. Therefore, in contrast to low temperatures, for Rochelle salt at room temperature the predominant contribution to α is made by the temperature dependence of the coefficient A. Consequently in this case the derivative of the second order phase transition temperature with respect to uniaxial pressure is determined mainly by the elastic properties of the coefficient A, whereas the influence of the volume expansion coefficient can be neglected in the first approximation.

It should be mentioned that the signs of the $\alpha_A$ values for $A_b$ and $A_c$ are caused only by the very large negative value of $(\partial S_{23}/\partial T)/S_{23}$ for Rochelle salt.

Taking into account the fact that $A\alpha_A = (\partial A/\partial T)$, in the first approximation Eq. (6) gets the following form for Rochelle salt and for the similar cases:

$$dT/dp_i = T\Delta[V(\partial A_i/\partial T)]/\Delta C_P . \qquad (8)$$

It is interesting to note that the signs of the uniaxial pressures derivatives of the transition temperature in Rochelle salt at 23.4 °C coincide with the signs of $(\partial A_i/\partial T)$. In contrast to Eq.(6), Eq.(8) is suitable only for uniaxial pressures, because for hydrostatic pressure it gives $dT/dP = 0$.

4. For almost 40 years (see [11]), without a satisfactory proof, the Ehrenfest equation (9) and formula (10) have been used for uniaxial pressures $p_i$.

$$dT/dp_i = VT\Delta\alpha_i /\Delta C_P , \qquad (9)$$

$$dT/dP = dT/dp_a + dT/dp_b + dT/dp_c . \qquad (10)$$

In [6] it has been shown that near 0 K the slopes $dT/dp_i$ found from Eq.(9) and the exact values may not only have different values, but also be of different signs. At the same time the absolute value of $\Sigma(dT/dp_i)$ can be more than three times larger than $dT/dP$. Such an error must be always present, when $\alpha_V$ changes the sign at transition. The formulae (9) and (10) are incorrect in general cases not only near 0 K, but at higher temperatures. Indeed, as in [6], let us rewrite Eq.(5) for uniaxial pressures $p_a$, $p_b$, $p_c$ and add them

$$\Sigma dT/p_i = T(V_1\alpha_1\Sigma A_{1i} – V_2\alpha_2\Sigma A_{2i})/(C_{P1} - C_{P2}) \qquad (11)$$

For Eq. (11) to coincide with Eq.(2) (i. e. $\alpha_{Ai} = 0$) it is necessary that $V_1 = V_2 = V$, $A_{1i}$, $A_{2i}$ and $\Sigma A_{1i}$, $\Sigma A_{2i}$ are equal to 1. But these conditions cannot be satisfied simultaneously, because even if $A_{1i} = 1$, $A_{2i} = 1$, then $\Sigma A_{1i} = 3$, $\Sigma A_{2i} = 3$. Besides, for uniaxial pressures all coefficients A are larger than 1.

5. As it was mentioned in [6] at the second order phase transitions under hydrostatic pressure the volume expansion coefficient $\alpha_V$, volume crystal compressibility ζ, and heat capacity $C_p$ have jumps. The jumps of the heat capacity and compressibility are of the same sign; see [12]. At the second order phase transition under uniaxial pressure, in addition to the afore-mentioned physical quantities, the linear compressibilities along the axes *a,b,c,* the coefficients $A_a$, $A_b$, $A_c$, and their temperature coefficients $\alpha_A$ also have jumps. The linear compressibilities along the axes *a, b, c* are equal to:

$$\zeta_a = - (\partial V/\partial p_a)/V = S_{11} + S_{12} + S_{13}$$
$$\zeta_b = - (\partial V/\partial p_b)/V = S_{22} + S_{12} + S_{23} \quad (12)$$
$$\zeta_c = - (\partial V/\partial p_c)/V = S_{33} + S_{13} + S_{23}$$

The volume crystal compressibility $\zeta$ is equal to the sum of the linear compressibilities along the axes *a, b, c*:

$$\zeta = - (\partial V/\partial P)/V = \zeta_a + \zeta_b + \zeta_c . \quad (13)$$

Since all linear compressibilities are larger than 0, the signs of the heat capacity $C_P$ jump and volume and linear compressibilities jumps coincide.

6. Now let us consider the second order phase transition under uniaxial pressure, assuming that the true volume is a continuous function of temperature (i. e. its jump is equal to zero):

$$\Delta V = 0 \quad (14)$$

By differentiating Eq.(14) with respect to temperature along the curve of transition points, i.e. assuming the uniaxial pressure $p_i$ to be the function of temperature given by this curve, we obtain:

$$\Delta(\partial V/\partial T)_{Pi} + (dp_i/dT)\Delta(\partial V/\partial p_i)_T = 0 \quad (15)$$

Along with the Ehrenfest equation (4) the dependence of the second order phase transition temperature on uniaxial pressure is described by the "second" Ehrenfest equation (16), which follows from Eq.(15)

$$dT/dp_i = \Delta\zeta_i/\Delta\alpha_V \quad (16)$$

For hydrostatic pressure an analogous differentiation of the Eq.(14) gives

$$dT/dP = \Delta\zeta/\Delta\alpha_V \quad (17)$$

Taking into account Eq.(13) we can always present the "second" Ehrenfest equation for hydrostatic pressure (17) as a sum of the derivatives of the second phase transition temperature with respect to uniaxial pressures applied along axes *a, b, c*:

$$dT/dP = dT/dp_a + dT/dp_b + dT/dp_c \quad (18)$$

Thus, we come to a contradiction: on the one hand, as it was shown earlier (Eqs.9,10), the derivative of the second order phase temperature with respect to hydrostatic pressure in general case cannot be presented as a sum of the three derivatives with respect to uniaxial pressures applied along axes *a,b,c*, but on the other hand, such a presentation is always possible (Eqs.17,18).

This contradiction is removed if the "effective" volume is the continuous function of the temperature:
$$\Delta V^* = \Delta(AV) = 0 \quad (19)$$

In this case by differentiation of the Eq.(19) for hydrostatic pressure we obtain Eq.(20) instead of Eq.(17)

$$dT/dP = (A_1V_1\zeta_1 - A_2V_2\zeta_2)/(A_1V_1\alpha_1 - A_2V_2\alpha_2). \tag{20}$$

This equation in no way can be presented as a sum of three terms corresponding to the uniaxial pressures applied along axes *a,b,c*. Therefore, the assumption that "effective" volume jump is equal to zero can be considered as proven. We hope that a direct experiment will confirm our conclusion in the nearest future.

7. Now let us consider the Ehrenfest equation for hydrostatic pressure (21):

$$dT/dP = TV\Delta\alpha_V/\Delta C_P \tag{21}$$

Taking into account Eq.(7), we can rewrite Eq.(21) as the sum of the linear ("uniaxial") parts of the derivative of the second order phase transition with respect to hydrostatic pressure (22),(23).

$$(dT/dP)_i = TV\Delta\alpha_i/\Delta C_p \tag{22}$$

$$dT/dP = (dT/dP)_a + (dT/dP)_b + (dT/dP)_c \tag{23}$$

The formula (22) differs from the formula (9) by the definition of the items: $(dT/dP)_i$ are the parts of the derivative with respect to *hydrostatic* pressure and have nothing in common with the derivatives with respect to uniaxial pressure $dT/dp_i$. By this reason the consideration of the Eq.(22) as determination of the $dT/dp_i$ and using them for comparison with the experimental values is inadmissible.

Eq.(23) is used successfully for investigation the features of the superconductors, see [1-2]. Unfortunately, the investigators name $(dT/dP)_i$ wrongly as "derivative with respect to uniaxial pressure". We hope that this misunderstanding will be eliminated.

# Conclusions

The Ehrenfest equation is derived rigorously for the case when the "effective" volume jump is equal to zero. The Ehrenfest equation for strained solids is rigorously generalized to an arbitrary temperature. It is shown that the second order phase transition temperature derivative with respect to uniaxial pressure depends on the temperature expansion coefficient of an "effective" volume. It is proved rigorously that at second order phase transition under uniaxial pressure the "effective" volume is the continuous function of temperature.

The Ehrenfest equation can be presented as a sum of three parts proportional to the crystal linear thermal expansion coefficients along axes *a,b,c*. These parts are equal to the second order phase temperature derivatives with respect to hydrostatic (but not uniaxial) pressure.